\documentclass[aps,prl,twocolumn,superscriptaddress,showpacs,longbibliography,lengthcheck]{revtex4-1}

\usepackage{amsmath,amssymb}
\usepackage[dvipdfmx]{graphicx}
\usepackage{color} 
\usepackage{txfonts}
\usepackage{xspace}  
\usepackage{multirow}
\usepackage{dcolumn} 

\usepackage{xspace}  
 
\newcommand{\etal}{\textit{et al.}\xspace}

\begin{document}


\title{Underlying structure of collective bands and self-organization in quantum systems} 

\newcommand{\ariken}{      \affiliation{RIKEN Nishina Center, 2-1 Hirosawa, Wako, Saitama 351-0198, Japan}}
\newcommand{\acns}{        \affiliation{Center for Nuclear Study, The University of Tokyo, 7-3-1 Hongo, Bunkyo, Tokyo 113-0033, Japan}}
\newcommand{\aut}{         \affiliation{Department of Physics, The University of Tokyo, 7-3-1 Hongo, Bunkyo, Tokyo 113-0033, Japan}}
\newcommand{\akul}{         \affiliation{KU Leuven, Instituut voor Kern- en Stralingsfysica, 3000 Leuven, Belgium}}

\newcommand{\aemp}{\email{otsuka@phys.s.u-tokyo.ac.jp}}  
 
\author{T.~Otsuka$^*$}     \aut \ariken \akul 
\author{Y.~Tsunoda}   \acns   
\author{T.~Abe}       \acns
\author{N.~Shimizu}    \acns 
\author{P.~Van Duppen}    \akul

\date{\today}

\begin{abstract}   
\bf{The underlying structure of low-lying collective bands of atomic nuclei is discussed from a novel perspective on the interplay between single-particle and collective degrees of freedom, by utilizing state-of-the-art configuration interaction calculations on heavy nuclei.
Besides the 
multipole components of the nucleon-nucleon interaction that drive collective modes forming those bands, 
the monopole component 
is shown to control the resistance against such modes.  The calculated structure of $^{154}$Sm corresponds to coexistence between prolate and triaxial shapes, while that of $^{166}$Er exhibits a deformed shape with a strong triaxial instability.
Both findings differ from traditional views based on $\beta$/$\gamma$ vibrations.  The formation of collective bands is shown to be facilitated from a self-organization mechanism.
} 
\end{abstract}

\maketitle

The structure of atomic nuclei exhibits single-particle as well as collective-mode aspects created by the protons and neutrons (nucleons).
The former has been characterized by the shell structure shown initially by Mayer
\cite{mayer1949} and Jensen \cite{haxel1949}, while the latter has presented 
a variety of nuclear shapes following Rainwater \cite{rainwater1950}, and Bohr and Mottelson \cite{bohr1952,bohr_mottelson1952,bohr_mottelson_book2}.  The two aspects lead to ``the problem of reconciling the simultaneous occurrence of single-particle and collective degrees of freedom ...''  \cite{bohr_mottelson_book1}.  
This is one of the most important basic questions in nuclear structure research, and it remains open.  For instance, G.E. Brown has addressed the question, ``how single particle states can coexist with collective modes'', throughout his life \cite{schaefer2014}.  We discuss in this Letter this problem from a novel perspective.

Nucleons in an atomic nucleus occupy single-particle orbits in various configurations.     
The effective nucleon-nucleon ($NN$) interaction in nuclei induces multi-nucleon correlations by mixing such configurations. This mixing occurs, in many cases, basically for ``valence'' nucleons in single-particle orbits on top of the appropriate closed proton and neutron shell (inert core).  
The ellipsoidal shapes correspond to such correlations having the nature of quadrupole surface deformation from a sphere, driven by the quadrupole component of the $NN$ interaction \cite{nilsson1955,elliott1958a,elliott1958b,bes1969,dufort1996,kaneko2011}. 
This gives rise to an interplay between collective mode and single-particle states (SPS's). 
If SPS's relevant to these correlations are separated by large gaps, the mixing between them and the resulting correlations are reduced.  Thus, the SPS's can act as a ``resistance'' to (the formation of) collective modes.  In this Letter, we first present how ellipsoidal shapes emerge from multi-nucleon systems by using the state-of-the-art Configuration Interaction (CI) simulations, called Monte Carlo Shell Model (MCSM)\cite{mcsm2001,shimizu2012}.   
This allowed us to uncover a novel mechanism: the monopole component of the $NN$ interaction shifts the single-particle energies (SPEs) effectively, weakening the ``resistance'' against deformation, and thus enlarging effects of the quadrupole interaction.  
For textbook examples of strongly deformed nuclei, such as $^{154}$Sm and $^{166}$Er, the obtained properties agree with experiments, but the underlying structures are shown to differ from the traditional interpretation \cite{bohr_mottelson_book2}. 
This mechanism can be interpreted as a quantum-mechanical self-organization\cite{selforg}.


\begin{figure}[bt]
  \centering
  \includegraphics[width=8.5cm]{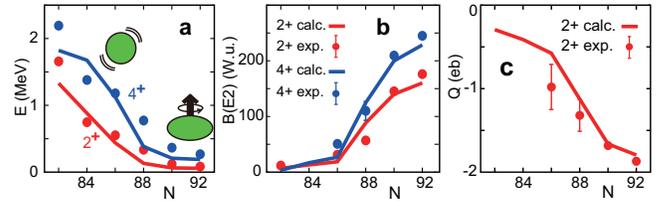}
  \caption{Systematic changes of the 2$^+_1$ and 4$^+_1$ levels in Sm isotopes, 
as functions of $N$.  {\bf a} Energy levels\cite{ensdf}, {\bf b}  
$B(E2;2^+_1 \rightarrow 0^+_1)$ and $B(E2;4^+_1 \rightarrow 2^+_1)$ values\cite{ensdf},   {\bf c.}  spectroscopic electric quadrupole moment of the 2$^+_1$ state\cite{stone_2005}.
  }   
  \label{fig:sm_sys}  
\end{figure}  

    We performed CI calculations on the samarium isotopes (proton number, $Z$=62) with even numbers of neutrons, $N$=82-92, and the $^{166}$Er ($Z$=68, $N$=98) nucleus,  without any assumptions of collective modes or shapes.  A many-body Schr\"odinger equation is solved for the input $NN$ interaction, which remains  the same for all calculations.  The valence proton (neutron) orbits are all orbits in the $sdg$($pfh$)-shell and the lower half of the next shell, implying the one-and-half harmonic oscillator shell on top of a $^{110}$Zr inert core.  This large model space is essential for the present study.
The CI calculations need the SPEs with respect to the inert core, the appropriate values of which are taken and kept; if available, based on known ones at $^{132}$Sn.   The effective $NN$ interaction is taken from the V$_{\mathrm{MU}}$ interaction for the proton-neutron channel \cite{otsuka2010} with a factor of 0.94 to its $T$=0 ($T$:isospin) central part.  The proton-proton and neutron-neutron channels are taken from  \cite{brownPb}.  For the cases where this recipe is not possible, the V$_{\mathrm{MU}}$ interaction is used.  
    Note that the V$_{\mathrm{MU}}$ interaction was determined as a simple modeling to the microscopic shell-model interactions \cite{otsuka2010}, and that it was used in earlier studies on Zr, Sn and Hg isotopes\cite{togashi2016,kremer2016,togashi2018,marsh2018,sels2019}.  
The dimension of the many-body Hilbert space reaches beyond 10$^{31}$, which is formidably larger than the current limit ($\sim$10$^{11}$) of the conventional CI method. 

Figure~\ref{fig:sm_sys}{\bf a} exhibits the experimental levels for Sm isotopes\cite{ensdf} as a function of $N$ in comparison to the presently calculated ones.  
The $E_x(2^+_1)$ is high for $N$=82, a magic number, but becomes lower steadily as $N$ increases.  The 2$^+_1$ states around $N$=86 imply a quadrupole surface oscillation on top of a spherical ground state.   
With larger $N$, the deformation becomes static out to an ellipsoid,  which generates a rotational band, a Nambu-Goldstone effect \cite{Nambu1960,Goldstone1961,Goldstone1962}. 
The $J^{\pi}$=4$^+_1$ levels exhibit a vibrational two-phonon pattern with $E_x(4^+_1)/E_x(2^+_1) \sim 2$ at $N$=86, while it evolves to the rotational value (10/3) for $N$=92, both in experiment and calculation\cite{casten_book}.

Figure~\ref{fig:sm_sys}{\bf b},{\bf c} display the $B(E2;2^+_1 \rightarrow 0^+_1)$ and $B(E2;4^+_1 \rightarrow 2^+_1)$ values 
and the spectroscopic electric quadrupole moment of the 2$^+_1$ state. 
The standard effective charges 1.5e (0.5e) for protons (neutrons)  are 
taken \cite{bohr_mottelson_book2}.  
An overall agreement between calculation and experiment suggests the validity of the present work, 
which is actually the first CI calculation for the shape evolution of samarium isotopes.  
Previously, this topic was studied in different frameworks (see {\it e.g.} \cite{Sm_X1,Sm_X2,Sm_X3,Sm_X4,Sm_X5}).  


\begin{figure}[bt]
  \centering
\includegraphics[width=7.5cm]{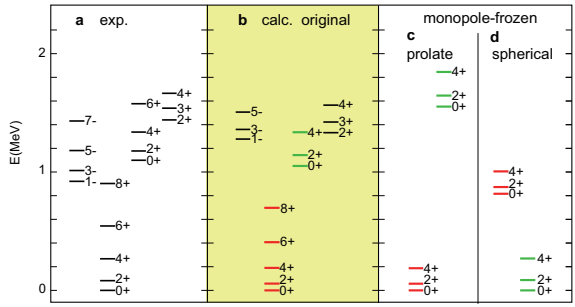}
\\
\vspace{1.0cm}
  \caption{Levels of $^{154}$Sm. 
  {\bf a} experimental levels\cite{ensdf}, {\bf b},{\bf c},{\bf d} calculated levels of the original and monopole-frozen Hamiltonians.  
  } 
\label{fig:154sm_level}
\end{figure}  

We now focus on $^{154}$Sm ($N$=92).  
Figure~\ref{fig:154sm_level}{\bf a},{\bf b} show its observed and calculated energy levels classified in four rotational bands\cite{ensdf}.
A good agreement between experiment and calculation is observed.
The band built on the ground state is prolate deformed, as confirmed by the large quadrupole moment and $B(E2)$ values shown in Fig.~\ref{fig:sm_sys}. 

\begin{figure}[!bt]
\begin{center} 
\includegraphics[width=8.2cm]{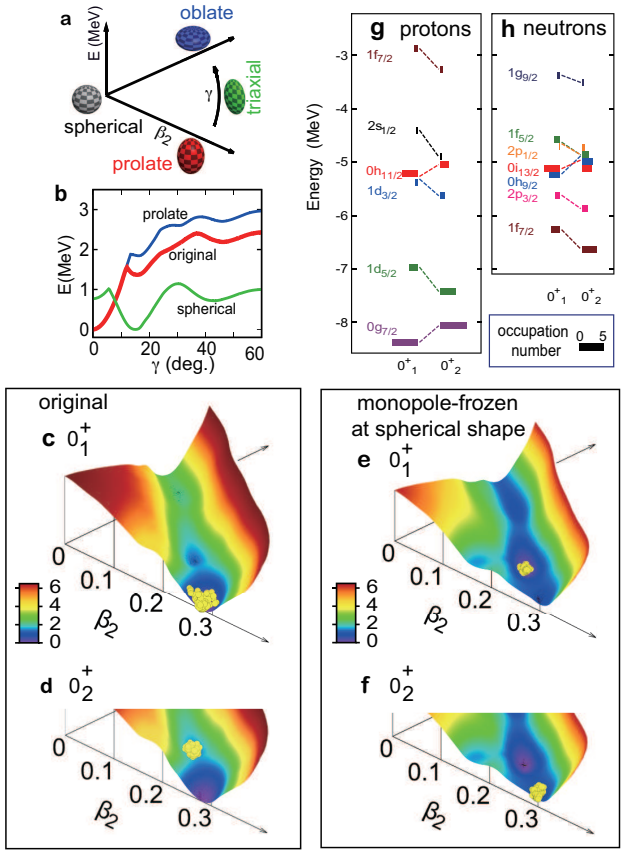}
  \caption{Properties of the 0$^+_{1,2}$ states of $^{154}$Sm. {\bf a} Deformation parameters and shapes. 
  {\bf b} lowest value of PES for a given $\gamma$ value for the original case and the prolate and spherical monopole-frozen cases.
{\bf c, d, e, f} 3-dimensional PES in the original or spherical monopole-frozen case (lower parts only in {\bf d} and {\bf f}).
{\bf g, h}  ESPE (vertical position) and occupation number (horizontal width).
}  
\label{fig:espe}
\end{center}
\end{figure}

Figure~\ref{fig:espe} shows a deeper insight in the calculation performed for $^{154}$Sm. It includes 
Potential Energy Surface (PES) obtained by Hartree-Fock calculation using the same Hamiltonian with, as constraints, the quadrupole moments corresponding to specific values of $\beta_2$ and $\gamma$ (their relations are explained in \cite{utsuno2015,marsh2018,sels2019}). The latter are shape variables of the ellipsoidal shape being the magnitude and the proportion of the ellipsoid axes, respectively (see Fig. 3{\bf a})\cite{bohr_mottelson_book2}.
Figure~\ref{fig:espe} includes panels of three-dimensional PES, relative to the lowest energy.  Figure~\ref{fig:espe}{\bf b} exhibits the lowest values of such PES values for a given value of $\gamma$.

We can now ``visualize'' the eigenstates obtained for $^{154}$Sm by using the so-called {\it T-plot}\cite{tsunoda2014,otsuka2016}.  An eigenstate is expanded by MCSM basis vectors.  One can assign partial coordinates to each MCSM basis vector by its $\beta_2$ and $\gamma$ values, and can plot it on the PES.  In this plot, the importance of each basis vector for this particular eigenstate is expressed by the size of the plotted circle.  
Figure~\ref{fig:espe}{\bf c} indicates that the T-plot circles for the ground state  
are concentrated around $\beta_2$ =0.28 and $\gamma$=0$^\circ$, a prolate shape.  The T-plots of 2$^+$-8$^+$ members of the band, independently obtained, exhibit very similar patterns.  This is interpreted as a strong signature of belonging to the same band.
Figure~\ref{fig:espe}{\bf d} shows T-plot circles belonging to the 0$^+_2$ state at 1.1 MeV.  The circles are concentrated in a local minimum at $\gamma\sim$20$^\circ$, a triaxial shape\cite{Davydov1958,Davydov1959}.
The 0$^+_{2}$, 2$^+_{2,3}$, 3$^+_1$ and 4$^+_{2,3}$ states show almost identical T-plots.  
Thus the present calculation indicates a 
coexistence between prolate and triaxial shapes, in a stark contrast to the conventional picture of the $\beta$/$\gamma$ vibrations (see {\it e.g.} \cite{krucken1999,casten_book,moller2012}).  
  
In order to clarify the underlying mechanisms giving rise to this picture, we 
decompose $NN$ interaction into two components: monopole and multipole interactions (see reviews {\it e.g.} \cite{poves1981,caurier_rmp,otsuka2016,otsuka_rmp}).  
The effect of the monopole interaction between protons in the orbit $j_p$ and neutrons in the orbit $j_n$ is expressed as  
\begin{equation}
v^m_{pn}(j_{p}, j_{n}) \, n_{p}(j_{p}) \, n_{n}(j_{n})
\label{eq:mono}
\end{equation}
where $v^m_{pn}$ is a coefficient called {\it monopole matrix element}, $n_p$($j_p$) denotes the number of protons in the orbit $j_p$, and $n_n$($j_n$) means likewise for neutrons.   
We first discuss how this interaction works.  
For the state being considered ({\it e.g.} the 0$^+_1$ state), $n_{n}(j_{n})$ takes integers, 0, 1, 2, ... with the corresponding probabilities.  Once one of these integers, $k$, substitutes $n_{n}(j_{n})$, this interaction energy becomes $k \, v^m_{pn}(j_{p}, j_{n})  \: n_{p}(j_{p})$, which 
represents a shift of the SPE of the orbit $j_p$ by $k \, v^m_{pn}(j_{p}, j_{n})$.  This has two important aspects, (i) the SPE is modified, (ii) the effect is proportional to the number of neutrons in $j_n$.  So, this varies effectively the SPE of the proton $j_p$ orbit depending which neutron orbits are occupied.    
By including contributions from all neutron orbits and also those due to the proton-proton interaction, we define the effective SPE (ESPE) of the proton orbit $j_p$.  Its expectation value is depicted in Fig.~\ref{fig:espe}{\bf g} for the 0$^+_{1,2}$ states.   The  expectation value of $n_{p}(j_{p})$ is shown also.  The same quantities for neutrons are displayed in panel {\bf h}.

If the monopole matrix elements were identical with respect to $j_p$ and  $j_n$ in eq.~(\ref{eq:mono}), the energy would be moved by the same amount for all states of a given nucleus.   
This is, however, not the case with the realistic $NN$ interaction: the finite range property of central force and the spin-isospin-dependence of tensor force produce significant variations with respect to $j_p$ and  $j_n$ in eq.~(\ref{eq:mono}) \cite{otsuka2005,otsuka2010,otsuka_rmp}.    

The multipole interaction contains various pieces, but the part most relevant to this work is  expressed as the coupling between the proton and neutron quadrupole moments (operators) as  $v^q_{pn} \, Q_p \, \cdot \,Q_n$,
where $v^q_{pn}$ is the interaction strength, and $Q_p$ and $Q_n$ denote, respectively, proton and neutron quadrupole moments coupled by a scalar product.  
With negative $v^q_{pn}$,  quadrupole moments produce binding energy through this interaction, which can result in strongly deformed states.  Namely, 
the quadrupole interaction thus defined is the driving force of the collective mode of quadrupole deformation.   
 
The quadrupole moment can become quite large, if several relevant single-particle orbits mix  coherently, as a realization of the Jahn-Teller effect \cite{jahn1937}.   If those orbits are far away from each other in energy, they cannot mix much.  
However, thanks to the monopole interaction and possible variations of the occupation pattern, the ESPEs can be shifted so that a more coherent mixing occurs giving more binding energy to the nucleus. 
This ESPE optimization does occur, and arises differently for each collective mode or shape, as visualized by the average ESPEs.  Figure~\ref{fig:espe}{\bf g,h} exhibit them for the $0^+_1$ and $0^+_2$ states of $^{154}$Sm, which are of prolate and triaxial shapes, respectively.       
Such differences are due to different occupation patterns (see horizontal levels in  Fig.~\ref{fig:espe}{\bf g,h}) and the afore-mentioned orbital dependences of the monopole interaction. 

In order to shed further light on this effect, we performed CI calculations without the monopole interaction and instead the average ESPEs are adopted as input SPEs ({\it i.e.} constants),  as denoted 'monopole-frozen' analysis.
Figure~\ref{fig:154sm_level}{\bf c} depicts levels by adopting the average ESPEs of the $0^+_1$ state.   The properties of ground-band members are rather unchanged from the original calculation naturally.  However, because the ESPE optimization for the $0^+_2$ state is completely ignored, the $0^+_2$ energy is raised by 50\%, as well as the $2^+_2$ and $4^+_2$ levels, as shown in Fig.~\ref{fig:154sm_level}{\bf c}.    
This is consistent with the PES: Fig.~\ref{fig:espe}{\bf b} indicates that the local minimum around $\gamma$=15$^\circ$ on the ``prolate'' line is higher by about 500 keV than the corresponding one of the ``original'' line.  
Another monopole-frozen analysis was made by taking average ESPEs calculated at the spherical limit of the PES.   
Figure~\ref{fig:espe}{\bf e}, {\bf f} display the PES and T-plot, indicating that the $0^+_1$ state is no longer prolate but triaxial, whereas the $0^+_2$ state becomes prolate.
Figure~\ref{fig:154sm_level}{\bf d} verifies this in terms of energy levels with the $0^+_1$ state being triaxial (green level) and the $0^+_2$ state being prolate (red level).  
These two examples of the monopole-frozen analysis demonstrate the crucial roles of the monopole interaction. 
 
The energy gain of the prolate minimum measured from the spherical limit is $\sim$8 MeV in the original calculation (see Fig.~\ref{fig:espe}{\bf c}). While it is only $\sim$5 MeV in the monopole-frozen analysis with the spherical limit (see Fig.~\ref{fig:espe}{\bf e}).   
Thus, the ESPE optimization is shown to lower the energy of the prolate state by $\sim$3 MeV.
The same analysis for the triaxial states indicates gain of about 1 MeV.   The ESPE optimization thus yields varying energy gains for the different shapes, lowering the prolate bands more.

Finally we note that also the presence of the negative-parity band built on the $1^-_1$ level has been reproduced (see Fig.~\ref{fig:154sm_level}{\bf a}), 
reinforcing the validity of the present calculation, particularly the shell gaps.  

\begin{figure}[bt]
\begin{center} 
\includegraphics[width=8.0cm]{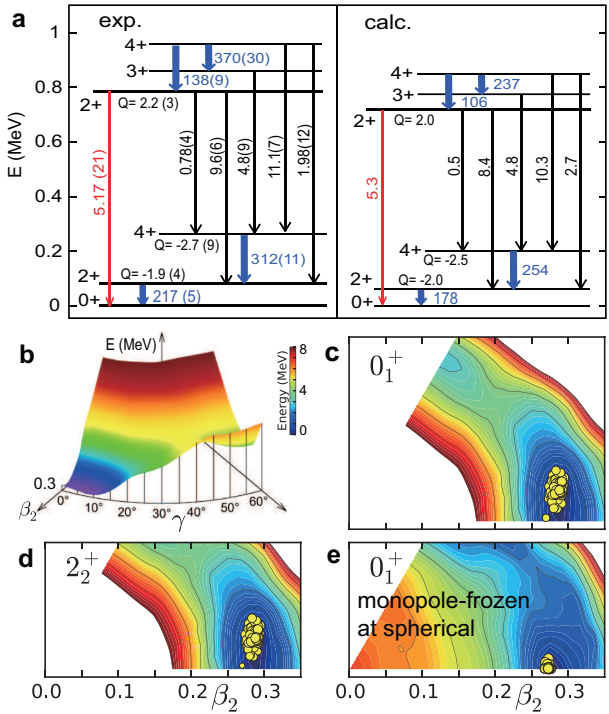}
      \caption{Experimental and calculated properties of the lowest states of $^{166}$Er. 
 {\bf a}  Energy levels and electromagnetic transitions (W.u.)\cite{ensdf} as well as spectroscopic electric quadrupole moments (eb)\cite{stone_2005}.  {\bf b} Three-dimensional PES and its cut surface for $\beta_2$=0.3.     {\bf c, d, e}  T-plots for the 0$^+_1$ and 2$^+_2$ states, and for the monopole-frozen 0$^+_1$ state at spherical shape.    
}   
\label{fig:166Er}
\end{center}
\end{figure}

The second example is $^{166}$Er, which has shape characteristics different from those of $^{154}$Sm.  The Hamiltonian is the same as that for the Sm isotopes except for a minor change for a better description: proton 1d$_{3/2}$ (0g$_{7/2}$) SPE shifted by 0.5 (-0.5) MeV.
Figure~\ref{fig:166Er}{\bf a} displays the lowest energy levels, calculated and measured \cite{ensdf}. The lowest side band starts from the $J^{\pi}$=2$^+_2$ level, called the $K^{\pi}$=2$^+$ band, which has been considered to be a $\gamma$ vibration, where two short axes of the elongated ellipsoid oscillate keeping the volume constant \cite{bohr_mottelson_book2,aage_bohr_nobel,ensdf}.  
One of the crucial quantities for this picture has been a relatively large value of 
$B(E2;2^+_2 \rightarrow 0^+_1)$, which is 5.17 (21) W.u. experimentally\cite{ensdf}.  This value is reproduced by the MCSM calculation to be 5.3 W.u.  The electric spectroscopic quadrupole moments of the 2$^+_{1,2}$ states are reproduced well\cite{stone_2005}.  
Thus, we have a salient description of the 0$^+_{1}$ and 2$^+_{1,2}$ states of $^{166}$Er as well as the other states (see Fig.~\ref{fig:166Er}{\bf a}).   

Figure~\ref{fig:166Er}{\bf b} displays the PES for $^{166}$Er, where 
the energy minimum is stretched in the $\gamma$ direction.  Figure~\ref{fig:166Er}{\bf c, d} display T-plot for the 0$^+_1$ and 2$^+_2$ state, which appear to be almost identical.  We thus see $\gamma$-softness with $\gamma$ between 5-15$^\circ$.  This does not seem to be compatible with the traditional picture of the prolate shape for the 0$^+_1$ state and the $\gamma$-vibrational excitation for the 2$^+_2$ state \cite{bohr_mottelson_book2}.  Rather, the T-plots indicate that these states stem from the same minimum in the triaxial plane.
The quadrupole invariant analysis \cite{kumar1971} made with the calculated quadrupole matrix elements of the states shown in Fig.~\ref{fig:166Er}{\bf a} results in 
$\gamma$=9$^\circ$ for the electric matrix elements and $\gamma$=10$^\circ$ for the mass ones, respectively, consistently with the above range of $\gamma$.  
In case of rigid triaxial deformation with $\gamma$=9$^\circ$, the 2$^+_2$ level is expected to be much higher in energy \cite{Davydov1959}, indicating that the present 
$\gamma$-softness lowers the band built on this state.
The monopole-frozen analysis using the spherical ESPE produces a completely different T-plot for the ground state with a concentration into the region of $\gamma <$4$^\circ$ 
(see Figure~\ref{fig:166Er}{\bf e}),
being consistent with a prolate shape.  Thus, the present ESPE optimization is crucial also for the $\gamma$-softness being discussed.  

The traditional interpretation of the ``band'' structures in $^{154}$Sm and $^{166}$Er is not supported by the MCSM calculations presented here. The underlying structure of $^{154}$Sm appears much more like a shape coexistence [44-47] between a prolate minimum and a triaxial minimum. While the possibility of another equilibrium was already mentioned by Bohr and Mottelson \cite{bohr_mottelson_book2}, it was not further investigated in detail (see also a 
recent review paper on the experimental findings\cite{Sharprey-Schafer2019}).
The present calculation indicated that by adding six protons and six neutrons, the two minima of $^{154}$Sm are moved closer and merged by reaching  $^{166}$Er, suggesting that nuclear forces can produce a wide diversity of structures.

Finally, we put the quadrupole-monopole interplay in the context of self-organization in atomic nuclei\cite{selforg}. 
In self-organization, a system is initially disordered, which corresponds to the SPEs without the optimization of the ESPE.  In the present work, order implies that the ESPEs are tailored to a specific shape.  Generally, some order may arise due to the self-organization in response to a change in external conditions.  
Atomic nuclei are, to a large extent, isolated quantum systems.  However, if there are two kinds of ingredients, like protons and neutrons, one can behave like the source of an external force on the other. 
By activating the present monopole mechanism, the ESPEs are organized so that more binding energies are gained, 
as compared to the SPEs without the optimization (``disordered" system).  
While these SPEs can be a resistance to the collective mode (as stated earlier), the monopole interaction can act as a resistance-control force; in the present context, it can bring a certain order to the system. 
While the final solution is determined self-consistently by including all components of the $NN$ interaction, 
a ``positive feedback'' can occur in this process especially between the quadrupole and monopole effects. 
As the present optimization effect varies for different shapes, it may appear to act purposely for a particular shape, for instance, a prolate one, although the monopole interaction as such has no connection to the deformation.  This ``to act purposely'' is one of the features of self-organization. 



In summary, CI calculations with a largest scale have clarified the structure of various bands in heavy nuclei.  The interplay between the quadrupole and monopole interaction produces significant effects, as a quantum version of the self-organization.
Namely, in a simplified view based on Landau's Fermi liquid theory \cite{landau1957}, nucleons are like free particles (quasiparticles) in the mean potential with fixed SPEs, and interact only weakly through a (residual) $NN$ interaction.  However, the actual structure seems to lie beyond this scope, because of the richness of the nuclear forces.   
The resistance-control force, {\it i.e.} the monopole interaction, does not promote the collective mode by itself, but can change the ``environment'' :
the ESPEs are optimized to different collective modes
giving rise to a large diversity of phenomena \cite{tsunoda2014,otsuka2016,leoni2017,morales2017,togashi2016,kremer2016,marsh2018,sels2019}.     
Although the CI calculations can be refined, the overall agreement to experiments suggests the validity of the underlying picture, shifting our basic understanding from the Liquid Drop model including its quantized forms to a more explicit multi-nucleon description with aspects such as shape coexistence, triaxial instability and shell evolution.     
This mechanism is expected to be more important in heavier nuclei including the superheavy ones, because more valence orbits and more nucleons imply more degrees of freedom for optimizing the SPEs. The relevance to superdeformation and clustering is of interest also, as particular configurations can be important. 
In heavy nuclei like uranium, new local minima or widening of minima in the $\gamma$ direction may appear at larger deformations on the PES, providing a new scope for the fission process, where the time evolution proceeds through possible tunneling paths linking those minima. 
It will be interesting also to explore similar self-organization mechanisms in other quantum systems as they can give rise to unexpected features.
 

%
\section{Acknowledgements}
T.O., Y.T., T.A. and N.S. acknowledge the support from MEXT as "Priority Issue on post-K computer" (Elucidation of the Fundamental Laws and Evolution of the Universe) (hp160211, hp170230, hp180179, hp190160) and JICFuS.
T.O. acknowledges Prof. N. Nagaosa for valuable comments.
T.O. and Y.T. acknowledge Prof. B. A. Brown for helping in handling interaction files.
This work has received funding from the Research Foundation Flanders (FWO, Belgium), the Excellence of Science program (EOS, FWO-FNRS, Belgium) and the GOA/2015/010 (BOF KU Leuven).

\end{document}